\documentclass[aps,prb,twocolumn,balance]{revtex4} 
\usepackage{natbib}
\usepackage{epsfig}
\usepackage{graphicx}
\usepackage{amsmath}
\usepackage{amsfonts}
\usepackage{amssymb}
\usepackage{cancel}
\usepackage{graphicx}
\usepackage{ulem}
\usepackage{wrapfig}
\usepackage{epsfig}
\usepackage{graphicx}
\usepackage{amsmath}
\usepackage{amsfonts}
\usepackage{amssymb}
\usepackage{graphicx}
\usepackage{wrapfig}
\usepackage{textcomp}
\usepackage{filecontents}
\usepackage{pgf,pgfarrows,pgfnodes,pgfautomata,pgfheaps,pgfshade}
\usepackage[colorlinks=true, linkcolor=blue]{hyperref}
\usepackage{lipsum}
\usepackage{balance}
\usepackage[utf8]{inputenc}  
\usepackage{blindtext}
\usepackage{textgreek}

\begin{document}

\title{Microwave response of a chiral Majorana interferometer}

\author{Dmitriy S. Shapiro$^{1,2,3 }$} \email{shapiro.dima@gmail.com} 
\author{Alexander D. Mirlin$^{4,5,6,7}$} \author{Alexander Shnirman$^{4,5}$}
\affiliation{$^1$Dukhov Research Institute of Automatics (VNIIA),  Moscow 127055, Russia}
\affiliation{$^2$Department of Physics, National Research University Higher School of Economics, Moscow 101000, Russia}
\affiliation{$^3$V. A. Kotel'nikov Institute of Radio Engineering and Electronics, Russian Academy of Sciences, Moscow 125009, Russia}
\affiliation{$^4$Institut f\"ur Theorie der Kondensierten Materie, Karlsruhe Institute of Technology, 76128 Karlsruhe, Germany}
\affiliation{$^5$Institute for Quantum Materials and Technologies (IQMT), Karlsruhe Institute of Technology, 76021 Karlsruhe, Germany}
\affiliation{$^6$Petersburg Nuclear Physics Institute,  St.Petersburg 188300, Russia}
\affiliation{$^7$L. D. Landau Institute for Theoretical Physics, Semenova 1-a, 142432 Chernogolovka, Russia}

\begin{abstract}

We consider an interferometer based on artificially induced 
topological superconductivity and chiral 1D Majorana fermions. 
The (non-topological) superconducting island inducing the superconducting 
correlations in the topological substrate is assumed to be floating. This allows probing 
the physics of interfering Majorana modes via microwave 
response,  i.e., the frequency dependent impedance between the island 
and the earth.  Namely, charging and discharging of the 
island   is  controlled by the time-delayed interference of chiral Majorana 
excitations in both normal and Andreev channels.   We argue that microwave 
measurements provide a direct way to observe the physics of 1D chiral 
Majorana modes.

\end{abstract}

\maketitle

\section{  Introduction} 
Physics of artificial topological superconductors with chiral Majorana edge modes 
was a subject of intensive research during the last decade~\cite{RevModPhys.83.1057, alicea2012,Beenakker-Rev,Kallin_2016}. Initially, these systems were proposed in hybrid structures on surfaces of topological insulators, covered by regular superconductors 
and magnetic insulators~\cite{FuKanePRL2008}. Later on, heterostructures based 
on quantum anomalous Hall insulators (QAHI) combined with regular superconductors~\cite{HeScience2017,shen2018spectroscopic} were experimentally 
studied. 
However, the reported evidence of the chiral Majorana fermions as half-quantized 
plateau in the two-terminal conductance~\cite{HeScience2017} is under debate~\cite{Kayyalha64}.
Further experimental advances were made in magnetic domains covered by 
superconducting monolayers~\cite{menard2017two} and in similar in spirit 
van der Waals heterostructures~\cite{Kezilebieke:2020ab} (see also a theoretical proposal~\cite{li2016two}). Surfaces of 
iron-based superconductors~\cite{Wang104} showed signs of topological superconductivity. Alternative realization of 1D 
Majorana edges in magnetic materials showing a spin liquid phase was reported in ~\cite{kasahara2018majorana}. 

Interferometers based on chiral Majorana modes should allow probing the 
nontrivial physics of these systems~\cite{FuKanePRL2009,PhysRevLett.102.216404,STRUBI2015489,PhysRevLett.107.136403,PhysRevB.85.125440,PhysRevB.93.155411,PhysRevB.95.195425,
PhysRevB.98.245405,PhysRevB.83.100512,PhysRevB.83.220510,PhysRevB.88.075304}. The first proposals addressed the 
dc-transport~\cite{FuKanePRL2009,PhysRevLett.102.216404}. Later, in a series of works 
the noise,   braiding of Majorana edge vortices, and time resolved transport were studied~\cite{Lian10938, PhysRevLett.122.146803, PhysRevB.102.045431, beenakker2020shot, AdagideliSciPost2019}.

Usually the regular superconductor, which induces the superconducting correlations 
in the topological material, is considered to have a fixed electrochemical potential.
We, in contrast, consider a floating island. This allows us investigating the 
time resolved charging and discharging dynamics of the island, which can be 
measured using microwave experimental techniques.

\section{    Qualitative picture } 
\begin{figure}[h!]
\includegraphics[width=0.73\linewidth]{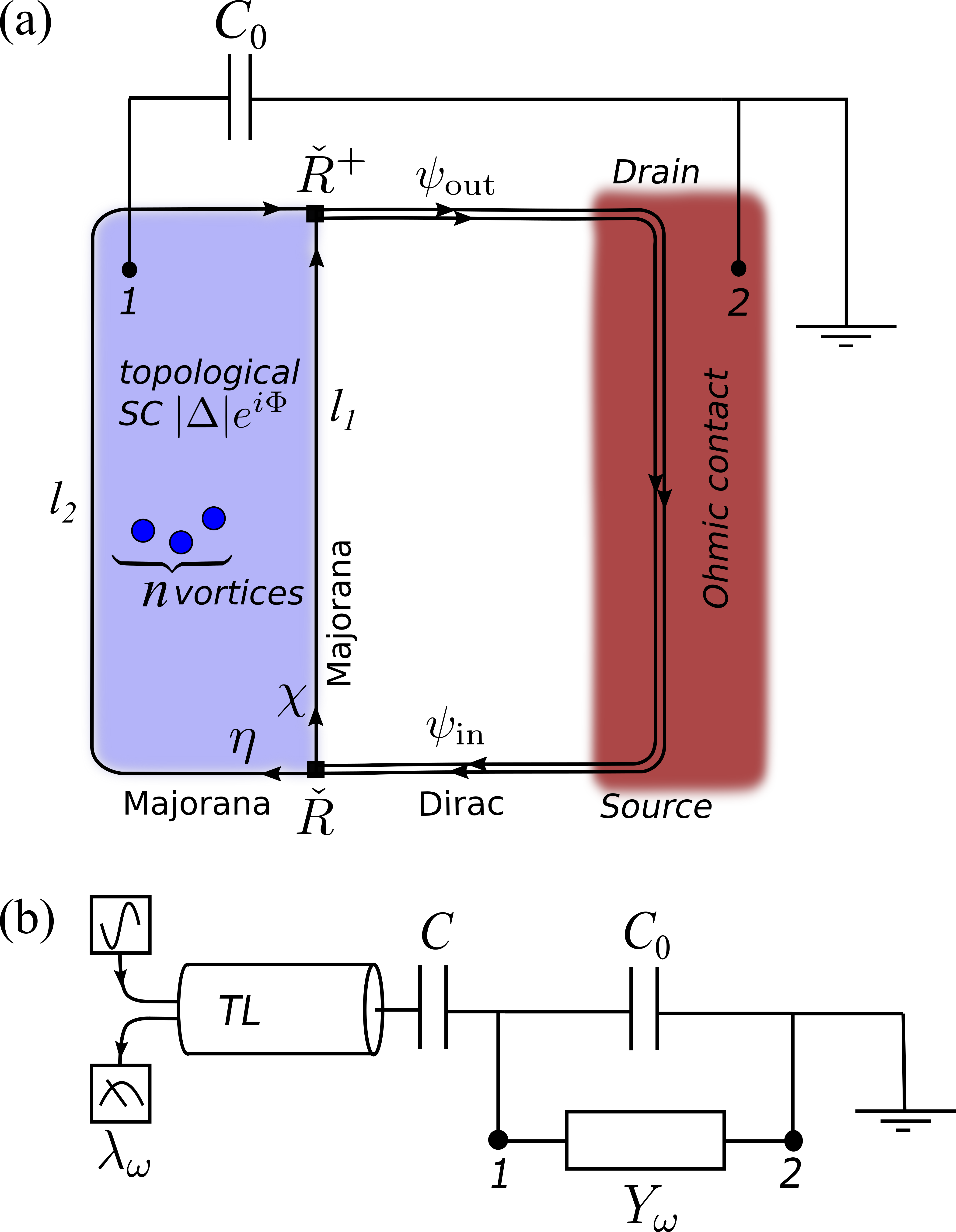} 
\caption{(a) Schematic description of the Majorana interferometer 
with a floating superconducting island. 
 Charge fluctuations on the island are allowed due to the finite self-capacitance $C_0\neq \infty$.    
Incident Dirac mode  $\psi_{\rm in}$ (doubled line) with equilibrium distribution function, imposed by the Ohmic contact, is  scattered into a pair of Majorana modes ($\chi$ and $\eta$, single lines).  $\chi$ and $\eta$ coherently propagate along the edges of lengths $l_1$ and $l_2$ and are fused back into $\psi_{\rm out}$. 
Scattering matrices of the lower  and upper Y-splittings (black bars) are  denoted by $\check R$ and $\check R^+$. Vortices in the superconductor induce an additional phase difference $n \pi $ between the interfering Majorana modes. (b) 
Equivalent electric circuit describing the interferometer in the linear response regime
with a transmission line attached for measuring the microwave response. 
Points 1 and 2 correspond to to points 1 and 2 in the subfigure (a)).
The input signal of a frequency $\omega$ is sent to a transmission line (TL) and a complex reflection coefficient $\lambda_\omega$ is measured.  
TL is coupled to the superconductor through the coupling capacitor $C$. 
}
\label{fig-interferometer}  
\end{figure}

We consider a system depicted symbolically in Fig.~\ref{fig-interferometer}(a).
Here, a single Ohmic contact serves as a source and a drain of chiral Dirac modes.
As the chiral Dirac mode approaches the superconducting area it is split into two 
chiral Majorana modes. The latter recombine later again into a chiral Dirac mode.
We assume the lengths of the two Majorana branches, $l_1$ and $l_2$, to be different, thus two different propagation times, $\tau_1=l_1/v$ and $\tau_2=l_2/v$ ($v$ is a Fermi velocity of surface states in a topological material). 
These time intervals determine the Thouless energy, $ E_{\rm Th}\equiv \frac{\hbar  }{\tau_1+\tau_2} $, and another energy, $\Lambda\equiv \frac{\hbar }{| \tau_1-\tau_2 |}  \ge E_{\rm Th}$. The superconducting island is floating and is characterized by a self-capacitance $C_0$ or, equivalently, by the charging energy $E_{\rm c}=\frac{e^2}{2C_0}$.

Below, using the effective action technique we derive the admittance $Y_{   \omega}$ (inverse impedance) of the island relative to the ground (source and drain), which is due to the currents in the edge modes.
That is,  $Y_{   \omega}$ corresponds to the admittance 
between points $1$ and $2$ in Fig.~\ref{fig-interferometer} (a) 
without the self-capacitance $C_0$. The total admittance between 
points $1$ and $2$ in Fig.~\ref{fig-interferometer}(a) is a sum of $Y_{\omega}$
and that due to $C_0$, i.e., $Y_{\omega} - i\omega C_0$. We obtain
\begin{equation}\label{eq:Admittance}
Y_{   \omega} 
/G_0 = 1 + 
 \frac{(-1)^n \pi T}{\sinh\left[ \pi T  \frac{(l_2-l_1)}{ \hbar v} \right] } \frac{e^{i \frac{l_1}{v} \omega }-e^{i \frac{l_2}{v} \omega }}{i\omega}\ , 
\end{equation}
where $G_0=e^2/(2\pi\hbar)$ is the conductance quantum, $k_{\rm B}$ is Boltzmann constant, $T$ is the temperature of a Fermi liquid in the Ohmic contact, and $n$ is a number of vortices in the superconducting island. In what follows, we set $k_{\rm B}=\hbar=1$ and restore them in final expressions.

\begin{figure}
\includegraphics[width=0.98\linewidth]{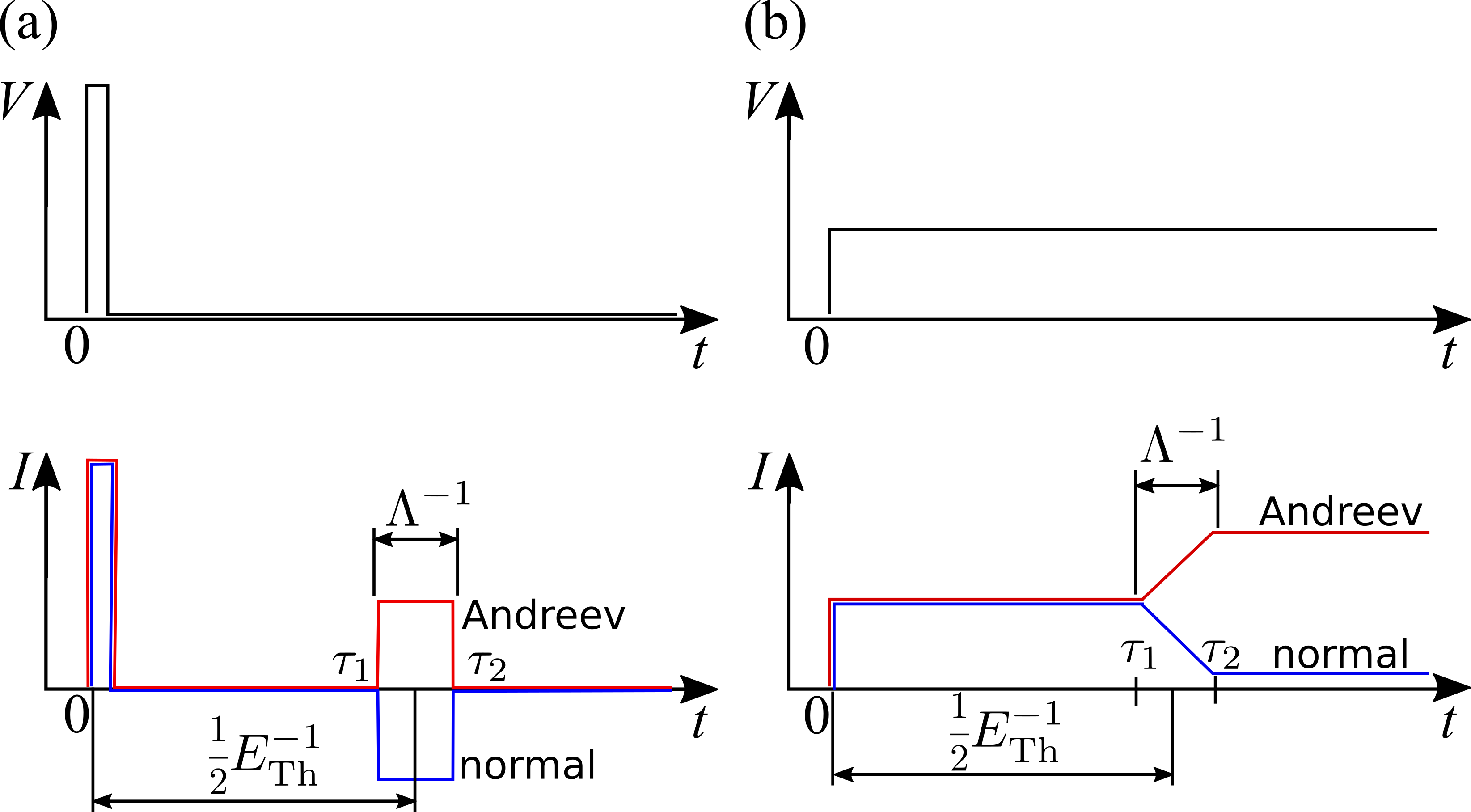} 
\caption{
{Response of the current $I(t)$ to voltage pulses $V(t)$ applied to the superconducting island of the interferometer.  (a) The delta-function-like pulse of $V(t)$ (upper panel) induces two response signals (lower panel): an instantaneous delta-pulse and a delayed step-pulse at $t\in  [\tau_1 ; \ \tau_2]$. If the integral over the voltage pulse  equals to the normal flux quantum, $V(t)= \frac{2\pi \hbar}{e}\delta(t)$, then the instantaneous peak in $I(t)$ transfers a $+e$ charge (at any temperature) and the 
delayed pulse -- a decreased  charge $\pm \alpha_T e $ in the Andreev (normal) 
reflection regime. For a voltage pulse with an integral corresponding to the superconducting flux quantum the transferred charges are halved, as discussed in Ref.~\cite{AdagideliSciPost2019}. Note that voltage pulses of such weights 
go beyond the linear response considered here.
(b) A step-like $V(t)$ (upper panel) induces an instantaneous step-like response with the conductance $G=G_0$ and a delayed response at $t\in [\tau_1 ; \ \tau_2]$. In the latter  $G$ interpolates linearly from $G_0$ to $G_0\pm \alpha_T G_0$ in the Andreev (normal) reflection regime.}}
\label{fig-response}  
\end{figure}

To understand the physical meaning of the admittance $Y_{\omega}$
it is useful to plot the response $I(t)$ of the current flowing into the island to a voltage 
pulse $V(t)$ applied to the contact ${   1}$ of Fig.~\ref{fig-interferometer}(a). This is, 
of course, given by $I(t) = \int dt_1 Y(t-t_1) V(t_1)$, where $Y(t)$ is the Fourier image of 
$Y_{\omega}$. The responses to a delta-like and a 
step-like pulses at $T=0$ are depicted in Figs.~\ref{fig-response}(a) and \ref{fig-response}(b),  
respectively.
The instantaneous response provided by the first term in Eq.~(\ref{eq:Admittance})
is explained by the immediate adjustment of the current in 
the outgoing Dirac mode to the new electrochemical potential of the island. This response corresponds to the effective conductance $G_0$. The delayed response is due to the interference of the Majorana excitations created by the voltage pulse at $t=0$. A similar effect (beyond the linear response analyzed here) was considered in Ref.~\cite{AdagideliSciPost2019}. The delayed response corresponds either to the normal or the Andreev reflection depending on the number of vortices in the island. In this chiral interferometer, the Andreev and normal reflections occur as a forward scattering from the incident into the outgoing Dirac channel. These are non-local  in space and time  processes with the amplitudes determined by the phases acquired by chiral Majorana excitations. Due to the spin texture of the Majorana modes 
a relative Berry phase $\pi$ is acquired in addition to the relative topological phase $n \pi$ due to $n$ vortices in the superconductor. Hence, the Andreev reflection regime is associated with odd $n=2k+1$ and the normal with even $n=2 k$,   $k \in \mathbb{Z}$.   The response to the delta-functional   pulse  coincides with  $Y(t)$ which  is nothing but the interferometer  Green function. A response to an arbitrary pulse is given by a convolution with $Y(t)$.

 We note that the floating  phase can be gauged out from the superconductor into Dirac modes.  
 We note a similarity with  the description of the transport in terms of wavepackets induced by voltage pulses (sometimes called ``levitons"~\cite{PhysRevLett.97.116403}). Namely, the immediate singular response in Fig.~\ref{fig-response}~(a) corresponds to an emission of a ``leviton" into the outgoing chiral channel. The delayed one is a transfer of another ``leviton'' through Majorana edge modes.

At zero temperature, for the step-like voltage pulse the current finally stabilizes at the value corresponding to conductance $G=2G_0$ in the Andreev reflection regime or $G=0$ in the normal reflection regime (see Fig.~\ref{fig-response}(b)). At finite $T$, the conductance  saturates at the values 
attenuated by thermal fluctuations, $G_\pm=G_0\pm  {\alpha_T} G_0$, with 
$\alpha_T=\frac{\pi T}{\Lambda\sinh\frac{\pi T}{\Lambda}}$.
At high temperatures, $T\gg\Lambda$, we obtain $G_\pm \rightarrow G_0$, 
which corresponds to a completely suppressed interference between the two Majorana branches.

The ac response $Y_\omega$ depends on both  $E_{\rm Th}$ and $\Lambda$, whereas, the dc response calculated in Refs.\cite{PhysRevLett.102.216404,FuKanePRL2009} does not involve $E_{\rm Th}$. 
We note that the admittance  $Y_\omega$ calculated here between points $1$ and $2$ assumes that the source and drain are grounded (see Fig. \ref{fig-interferometer}(a)). 
In Refs.\cite{PhysRevLett.102.216404,FuKanePRL2009}, the  dc conductance was calculated in the alternative setting where the drain and the superconductor were  grounded. The zero frequency limit for the admittance, $Y_{\omega=0}$, reproduces the results of those works for dc conductances  
in the linear response limit.

\section{    Proposed measurement} 
\begin{figure}[h!]
\includegraphics[width=1\linewidth]{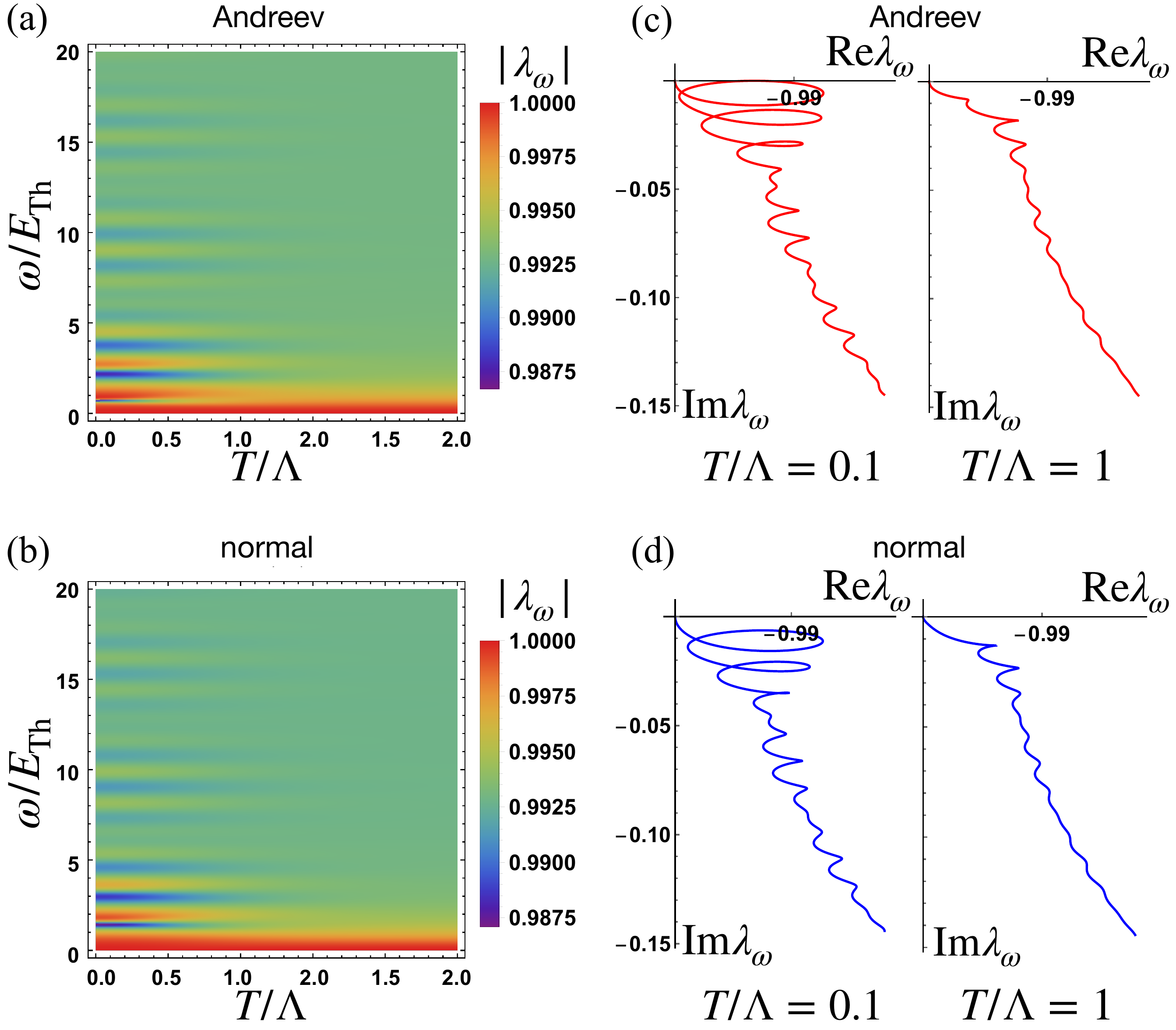} 
\caption{
Reflection coefficient $\lambda_\omega$ as a function frequency $\omega$ and temperature $T$. (a) $|\lambda_\omega |$ in the Andreev reflection regime with odd $n$ and (b) $|\lambda_\omega |$ in the normal reflection regime with even $n$. $| \lambda_\omega |$ oscillates as a function of $\omega$ with two sub-periods given by $\Lambda$ and $E_{\rm Th}$. The amplitude of oscillations decays exponentially with $T$. The deviation of  $| \lambda_\omega |$ from unity is of order $\sim 1 \%$.  Data shown for asymmetric device $l_2/l_1=4/3$ and $\Lambda/E_{\rm Th}=7$. The charging energy is equal to the level spacing, which means  $E_c=2\pi E_{\rm Th}$, or $C_0=\frac{G_0}{v}\frac{l_1+l_2}{2}$. We also assume $C=C_0$.
(c) and (d) Parametric plots of ${\rm Re}  \lambda_\omega $ and ${\rm Im}  \lambda_\omega$ for $\omega\in [0 , 20 E_{\rm Th}]$.}
\label{fig-lambda}  
\end{figure}

We propose to couple the superconducting island to a microwave waveguide (transmission line), as shown 
in Fig.~\ref{fig-interferometer}(b). 
One should be able to measure the reflection amplitude given by
\begin{equation}
\lambda_\omega=-1+ \frac{2 Z_{\rm TL}}{Z_{\rm TL}+1/(-i\omega C)+1/(Y_\omega-
i \omega C_0)} \ . \label{lambda}
\end{equation}
Here, $Z_{TL}$ is a transmission line impedance, which in an idealized situation approaches  $Z_0\approx 376.7$Ohm, the free space impedance. The experimentally relevant capacitances should be of the same order,   $C\sim C_0$.  The reflection coefficient is close to $\lambda_\omega=-1$. This means that  the measured response is determined by the fine structure constant,  $\alpha=Z_0 G_0\approx \frac{1}{137}$,   i.e., the effect is of the order of $1\%$.
The     function $\lambda_\omega$ shows decaying oscillations as a function of $\omega$ with periods proportional to $E_{\rm Th}$ and $\Lambda$ (see Figs.~\ref{fig-lambda}(a) and (b)). 
 We mention that contemporary experimental methods allow to increase $Z_0$ up to the resistance quantum, $1/G_0$, and even higher. This is possible in superinductors  realized as ladders of  Josephson junctions~\cite{Manucharyan113,PhysRevLett.109.137003} and   high-kinetic-inductance materials~\cite{PhysRevLett.121.117001}. Thus, the effect can be enhanced  significantly.

	Equation  (\ref{lambda}) allows one to extract $Y_\omega$  from the   experimental data for $\lambda_\omega$ (and, thus, $Y(t)$). 	An observation of   the delayed response can be a conclusive evidence of interfering Majorana fermions.

An observation of the delayed  response  can be disrupted by quasi-particle poisoning and by fluctuations of vortices parity. On the other hand, the random charge fluctuations, which are usually the main  mechanism of the noise, do not pose any problem as they decouple in the linear response regime.

\begin{figure}
\includegraphics[width=0.79\linewidth]{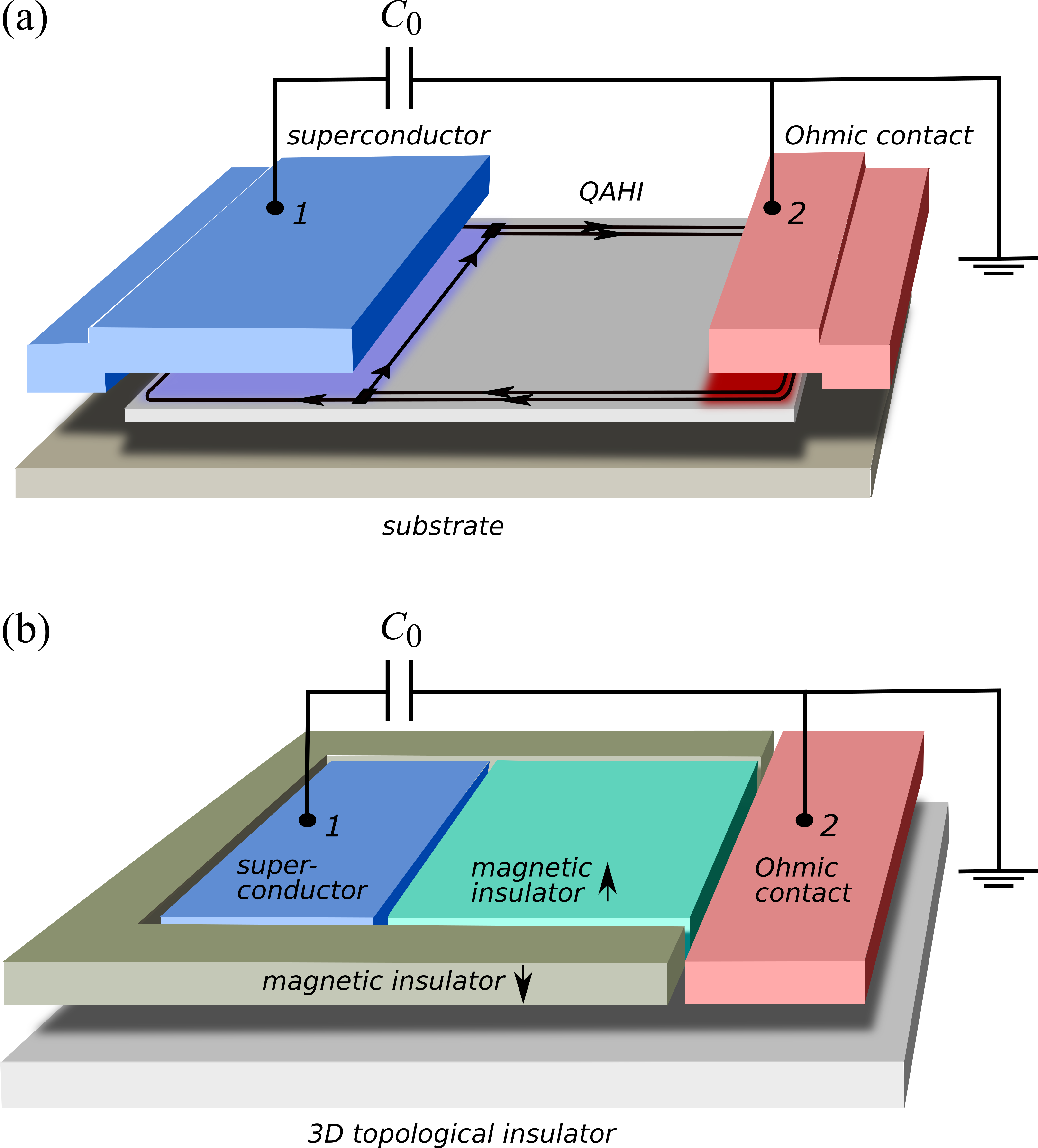}  
\caption{
{
   Possible physical realizations of the interferometer.
	  (a) A heterostructure composed of QAHI and a superconducting island. Incident Dirac modes (double lines) split at the Y-splittings (black bars) into Majorana modes (single  lines), which surround   topological superconducting region (light purple shaded region).  (b)  The device is realized on top of a 3D topological insulator, which is covered by magnetic insulators with opposite magnetizations (up and down arrows) and a superconductor.  Chiral  Dirac fermions  propagate  along the magnetic domain walls  and convert into  Majorana modes at  the interface to the superconducting area. 
	 }
}
\label{fig-device}  
\end{figure}

Two possible physical realizations of the interferometer are depicted 
in Fig. \ref{fig-device}. The first realization is a QAHI film covered by a superconductor 
as shown in Fig. \ref{fig-device}(a).  
The second realization is a 3D topological insulator  covered with magnetic insulators of opposite magnetizations, and a superconductor (Fig. \ref{fig-device}(b)).

\section{   A sketch of the derivation} 
\subsection{    Limit of zero phase fluctuations: Scattering approach and effective action  } 
We first describe the mean-field situation in which the superconducting order 
parameter $\Delta$ does not fluctuate. Since we have just a single superconducting 
island we can assume $\Delta$ to be real. The Bogoliubov-de Gennes equations describing setups like those in Fig.~\ref{fig-device}(b) have been extensively discussed in the literature~\cite{FuKanePRL2008,FuKanePRL2009}. At low energies, only the Dirac or Majorana edge modes are relevant since the bulk is gapped everywhere.
For the Dirac edge mode, which emerges from the source, the
action reads $\mathcal{S}_{\rm in}[\psi]= \! \int \! dx_1 dt_1 dx_2 dt_2  
  \bar\psi_{\rm in}(x_1 ,\! t_1) {\mathbf{G}^{-1}(x_1\! , x_2 , t_1\! , t_2)}\psi_{\rm in}(x_2 ,\! t_2)$.
For the relevant values of $x_1$ and $x_2$ (between the source and the first Y-splitting) the inverse propagator ${\mathbf{G}}^{-1}$ is obtained by the Fourier transform of 
\begin{equation}
	\mathbf{G}_{k,\omega} ^{-1} = 
	\begin{bmatrix}
\omega{-}vk{+}i o (1{-}2n_k)	 & 
-2i o n_k	 	\\
 2io (1{-}n_k) & vk{-}\omega{+}i o (1{-}2n_k)   	\\
\end{bmatrix} \ .\label{Ginv-el}
\end{equation}
Here,    
$
n_k= 1/(1+\exp\frac{vk}{T})  
$
is the equilibrium distribution function dictated by the  Ohmic contact, and $o$ is an infinitesimal positive frequency. This is a matrix in the $+/-$ basis of the Keldysh space (Pauli matrices in this space are denoted by $\boldsymbol \sigma$)~\cite{Kamenev}. 
Introducing the Nambu spinor $\check\Psi_{\rm in}=[\psi_{\rm in},\bar\psi_{\rm in}]^T$ we rewrite 
the action as
$\mathcal{S}_{\rm in}[\check\Psi_{in}]= \frac{1}{2}\int  
  \check\Psi_{in}  
  \check \tau_x \check {\mathbf{G}}^{-1} 
  \check\Psi_{\rm in}
$, where 
$\check {\mathbf{G}}_{k,\omega}^{-1} =   \check\tau_+\check\tau_-\mathbf{G}_{k,\omega} ^{-1}   -  	\check\tau_-\check\tau_+ [{\mathbf{G}}_{- k, - \omega} ^{-1}]^T $ and $\check\tau_\pm=\frac{1}{2}(\check\tau_x\pm i \check\tau_y)$.
The Pauli matrices $\check\tau$ act in the Gor'kov-Nambu particle-hole space and 
$\tau_z=\pm 1$ correspond to electron-like or hole-like state in the Dirac channel.

The scattering matrix $\check R$ for  the lower (first) Y-splitting describes the conversion of an incident Dirac electron and hole into a pair of Majorana particles $\chi_{k}$ and $\eta_{k}$:
\begin{equation}
	 	\begin{bmatrix} \chi_{{\rm out}; k}  \\
		\eta_{{\rm out}; k}
			\end{bmatrix} = \check R
			\begin{bmatrix} 
			\psi_{{\rm in}; k}  \\
		\bar\psi_{{\rm in}; -k}
	\end{bmatrix}
	\ , \ \check R= \begin{bmatrix} \frac{1}{\sqrt 2} & \frac{1}{\sqrt 2} \\
		\frac{i}{\sqrt 2} & \frac{-i}{\sqrt 2}
	\end{bmatrix}
	\ . 
\end{equation}
The Hermitian conjugated $\check R^+$ of the   upper (second) Y-splitting describes the conversion of Majorana modes into outgoing Dirac fermions:
$
		\begin{bmatrix} 
			\psi_{{\rm out}; k}  \\
		\bar\psi_{{\rm out}; -k}
	\end{bmatrix} {=} \check R^+
	\begin{bmatrix} \chi_{{\rm in}; k}  \\
		\eta_{{\rm in}; k}
			\end{bmatrix}
$. 
Finally, a relation between in- and out- Dirac states   $\check\Psi_{{\rm in}; k} {=}[\psi_{{\rm in}; k}  \ \bar\psi_{{\rm in}; -k}]^T$ and $\check\Psi_{{\rm out}; k} {=}[\psi_{{\rm out}; k}   \ \bar\psi_{{\rm out}; -k}]^T$  reads 
 \begin{equation}
 \check\Psi_{{\rm out}; k} =  \check S_k \check\Psi_{{\rm in}; k} \ . 
 \end{equation}
  Here, the scattering matrix is found as $\check S_k =\check R^+ \check F_k \check R $ where $\check F_k={\rm diag}\{ (-1)^{n+1}e^{ikl_1},\ e^{ikl_2} \}$ determines Berry and topological phases, and dynamic phases $k l_{1,2}$ of coherently propagating Majorana excitations. 
  
With the help of the above introduced scattering matrix we now transform 
from the basis of incoming Dirac states to the basis of exact scattering states. 
That is, the field $\check\Psi_k$ now annihilates an exact scattering state in the whole setup. The action retains its form, i.e., 
\begin{equation}
\mathcal{S}_0[\check \Psi]= \frac{1}{2}\int  
  \check\Psi  
  \check \tau_x \check {\mathbf{G}}^{-1} 
  \check\Psi \ .
  \end{equation}
  
The scattering matrix allows us to represent the current flowing into the island, $I\!=\! ev( \bar \psi_{\rm in}\psi_{\rm in} \!\! -\! \bar\psi_{\rm out} \psi_{\rm out})$, as
\begin{equation}
	I[\check\Psi ]= \frac{1}{2}    
	\check\Psi_p  
	\check\tau_x \check J_{p,k}   \check\Psi_k 
	 \ ,  \label{I[Psi]}
\end{equation}
where the   matrix $\check J_{ p,k}= v\check \tau_z-v \check S_{p}^+  \check \tau_z  \check S_{k} $ has a  non-diagonal structure  in momentum and Gor'kov-Nambu spaces.

\subsection{ Regime of fluctuating phase: Gauge transform and derivation of the interaction part $\mathcal{S}_{\rm int}$ in the effective action }
Next we allow the phase $\Phi$ of the order parameter $\Delta = |\Delta|e^{i\Phi}$ to fluctuate.   The scattering of Dirac fermions   becomes inelastic in this case; the energy-dependent~\cite{BLANTER20001} scattering matrix obtains a non-stationary structure.  Below we derive an interaction term in the action which couples fermion and boson degrees of freedom.  We perform 
a standard gauge transformation of  fermion phases~\cite{Levitov1996,PhysRevLett.85.1294,Kamenev}, which makes the superconducting order parameter real, $|\Delta|e^{i\Phi} \to |\Delta|$. 
The Majorana edge modes are transformed accordingly and we extend 
the gauge transformation infinitesimally   into the incoming and outgoing Dirac modes.
That is 
\begin{equation}
\psi_{\rm in}(x,t)\to \exp\left[ \frac{i}{2}\theta(x- (z_1-\epsilon))\Phi(t)\right ]\psi_{\rm in}(x,t) \ , \label{g-transf-1}
\end{equation} where $z_1$ is the coordinate of the first Y-splittings and $\epsilon\to 0$ at the later stage of the derivation. Similarly 
\begin{equation}\psi_{\rm out}(x,t) \to \exp\left[ \frac{i}{2}\theta((z_2+\epsilon)-x)\Phi(t) \right]\psi_{\rm out}(x,t) \ , \label{g-transf-2}
\end{equation} 
where $z_2$ is the coordinate of the second Y-splitting. The discrete symmetry $\Phi\to \Phi + 4\pi$ is preserved.

\begin{figure} 
\center \includegraphics[width=0.7\linewidth]{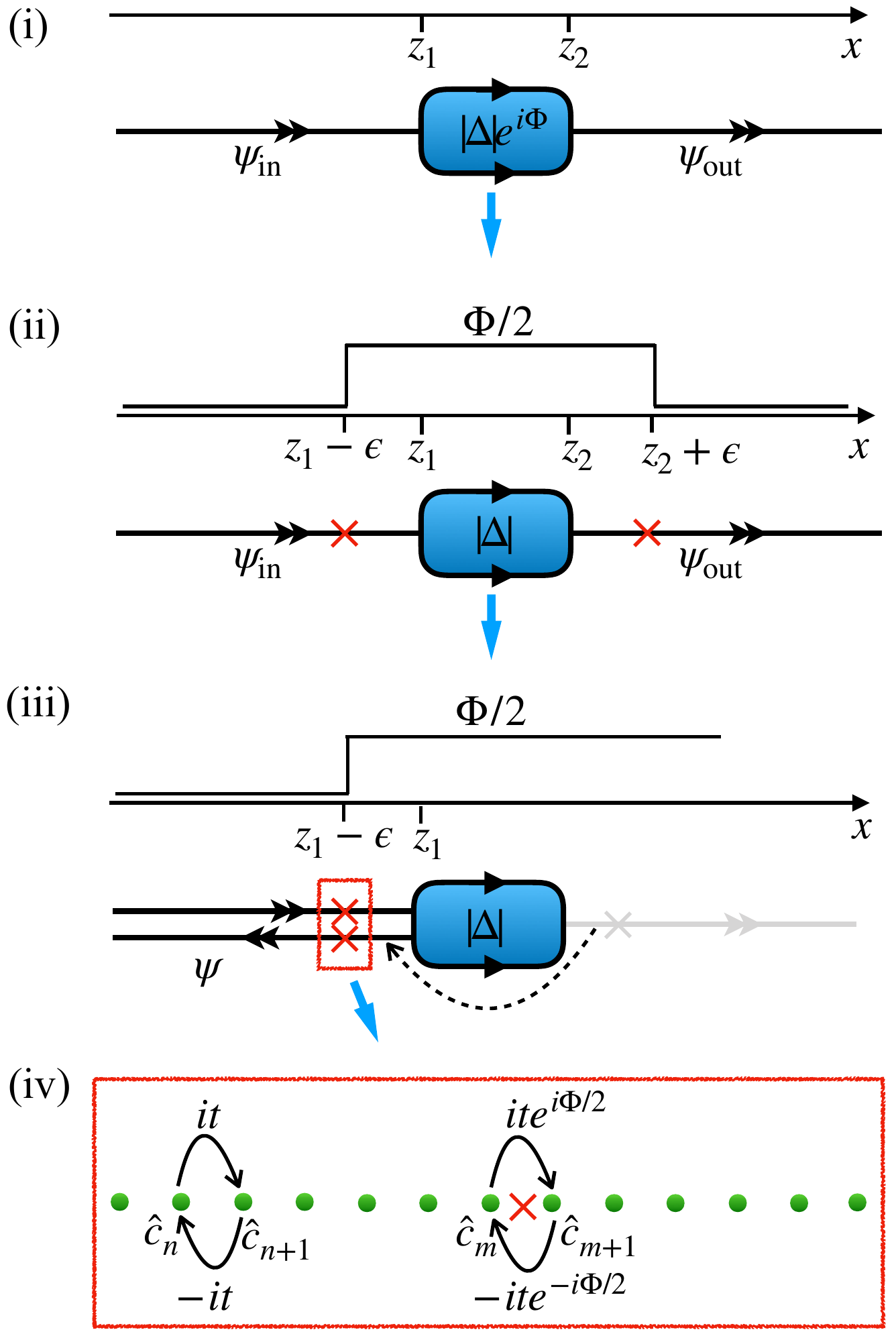}  
\caption{
Schematic derivation of $\mathcal{S}_{\rm int}$: (i) representation of chiral modes on a line, (ii) gauge transformation (\ref{g-transf-1}, \ref{g-transf-2}), (iii) mapping onto non-chiral mode, and (iv) mapping of the mode $\psi$ near the phase drop  onto a fermion lattice with $\hat H+\delta  \hat H$.
}
\label{gauge}  
\end{figure}

 
In the above discussed gauge transform, we encounter the  theta-function regularization problem because we go beyond the long wave-length approximation. We resolve it using  a discretized tight-binding approach.  
The derivation is based on four steps (i-iv) schematically illustrated in Fig.~\ref{gauge}.
  Let us  
  map  chiral Dirac modes $\psi_{\rm in}$ and $\psi_{\rm out}$  onto a line with the coordinate $x$ (Fig.~\ref{gauge} (i)). The Y-splittings are located at $x=z_{1,2}$, the incident mode $\psi_{\rm in}$ scatters into a pair of Majorana modes  surrounding the topological superconductor with floating phase of the order parameter, $|\Delta| e^{i\Phi}$, which in turn fuse into $\psi_{\rm out}$.
  
In the second step (see Fig.~\ref{gauge} (ii)), we  gauge out the superconducting phase, $|\Delta|e^{i\Phi}\to |\Delta|$,   in accordacne with Eqs. (\ref{g-transf-1}, \ref{g-transf-2}). As mentioned, we extend the  the  step-like  gauge transformation slightly beyond the superconductor, i.e., we perform it in the interval $z_1-\epsilon<x<z_2+\epsilon$. 
This  gives two phase drops at $x=z_1-\epsilon$ and $x=z_2+\epsilon$, which are marked by red crosses.  
  
In the third  step we transform the chiral modes $\psi_{\rm in  }$   and $\psi_{\rm   out}$ to a non-chiral mode $\psi(x)$ on a semi-axis (Fig.~\ref{gauge} (iii)). 
Both phase drops are now merged into one, which is marked by the red bar. In this representation the problem is similar to that considered in Ref.~\cite{PhysRevLett.97.116403}. In the final  step, the non-chiral $\psi$  near the phase drop  is mapped onto a 1D tight-binding lattice of fermions (Fig.~\ref{gauge} (iv)).  
    The corresponding tight-binding Hamiltonian for $\Phi=0$ reads
 \begin{equation}\hat H=\frac{it}{2}\sum_n  (\hat c_{n+1}^\dagger \hat c_{n}^{\phantom\dag} - \hat c_n^\dagger \hat c_{n+1}{\phantom\dag} ) \ . \label{H-tb}
 \end{equation} 
 It has a spectrum $\varepsilon=t \sin k a$ with  $k\in [-\frac{\pi}{a},\frac{\pi}{a}]$, $t$ is a hopping energy and $a$ is a lattice constant. In the low energy limit, $\varepsilon \ll t$, we obtain right (left) moving fermions near $k=0$ ($k=\pm\frac{\pi}{a}$). These states determine the long wave-length behavior of in- and out- chiral fermions in our setting.

Now we consider two sites, $n=m$ and $n=m+1$, where the phase drop $\Phi\neq 0$ occurs (it is shown by the red cross in Fig.~\ref{gauge} (iv)). The discrete version of the gauge transform (\ref{g-transf-1}, \ref{g-transf-2})  
corresponds to $\pm\Phi/2$ phase shifts accumulated when a fermion hops between these cites. The modification of the matrix elements describing the hopping between cites 
$m$ and $m+1$ , $t \to t e^{\pm i\Phi/2}$, in (\ref{H-tb}) results in $\delta  \hat H$ being added to the full Hamiltonian, $\hat H \to \hat H+\delta  \hat H$. Here
 \begin{equation}\delta \hat H{=}\frac{it}{2}\int\frac{dk dp}{(2\pi)^2}\hat c^\dagger_k\hat c_p[e^{-i pa }(e^{ i\frac{\Phi}{2}} {-}1) {-} e^{-i ka }(e^{- i\frac{\Phi}{2}} {-}1)] \ . 
 \end{equation} 
(Without loss of generality we set $m=-1$ in this expression.) The low energy limit  of  $\delta \hat H$  yields $\mathcal{S}_{\rm int}$ after identifying $ta \to v$,  $\hat c_k\to   \psi_{\rm in}$ at  $k a \ll 1$, and     $\hat c_k \to   \psi_{\rm out}$ at $ka\approx \pm\pi$.  After the transition to this long wavelength limit we send $\epsilon$ to zero.

As a result, we obtain an interaction term  in the action 
\begin{equation}\mathcal{S}_{\rm int}[\check\Psi,\Phi] =\int dt\left(  \frac{I[\check\Psi]}{e}  \sin\Phi/2 + \frac{U[\check\Psi]}{\hbar} (\cos\Phi/2-1) \right) \ . \label{S-int}
\end{equation}
Here 
$U[\check\Psi] = \hbar a v(\bar \psi_{\rm in}\psi'_{\rm in}+ \bar\psi_{\rm out} \psi'_{\rm out})$ with $a$ being the lattice constant. It is important to keep here the term $\propto U$, although one 
could be tempted to drop it in the continuous long wave-length limit. This term provides an important regularization in what follows.

\subsection{Quasi-classical Keldysh functional}
These steps lead to the following effective action on the Keldysh contour $\mathcal{C}$
\begin{equation}
\mathcal{S}[\check\Psi,\Phi]{=}\!\int_{\mathcal{C}}\!\left[ \frac{\hbar \dot\Phi^2 }{16E_{\rm c}}  { +}\frac{ 1}{2 e}    \Phi     I[\check\Psi] { -}\frac{\omega_c}{8\pi^2} \Phi^2 \right] dt + S_0[\check\Psi]  
 \ . \label{S-0}
\end{equation}
(We set $e=\hbar=k_{\rm B}=1$ below and restore them in final expressions.) 
The first term in (\ref{S-0}) is the usual charging energy, where we have employed the 
Josephson relation $V=\dot\Phi/2$, $V$ being the scalar potential.  We assume that the fluctuations of phase $\Phi$ are small (quasi-classical regime) because of large $C_0$. 
Thus we have expanded $S_{\rm int}$ up to a quadratic order in $\Phi$.  This yields the second term giving a linear coupling of the phase variable to the current fluctuations, and the third one, which plays a role of the diamagnetic counter term. It involves a divergent negative energy of the ground state $\langle U \rangle = - a v \frac{k_c^2}{2\pi}$.
The cutoff momentum is chosen as $k_c=\frac{2}{\pi a}$ such that the upper frequency cutoff in our theory, $\omega_c=v k_c$, and we obtain $\langle U \rangle=\omega_c/\pi^2$.  

Integrating over  $\Psi$ we obtain the effective action for the phase $\mathcal{S}[\Phi] =-i \ln [ \int D[\check\Psi] \exp(i\mathcal{S}[\check\Psi,\Phi])]$. It reads
 \begin{multline}
	\mathcal{S}[\Phi]  =   
	\int \Phi_{\rm q}(t)\left(  - \frac{\omega_c}{ 4 \pi^2}-\frac{1}{ 
		8E_{\rm c}}\partial_t^2\right)\Phi_{\rm cl} (t)dt  - \\
		{-}  i\frac{1}{2}{\rm Tr  \ln}  
	\Big[ \check{\mathbf{1}}   { +}  \frac{1}{2}\check{\mathbf {G}}_{p}(t{-}t')   \check J_{ p,k}  \Big( \boldsymbol \sigma_z\Phi_{\rm cl}(t'){+}{\boldsymbol \sigma_0}\Phi_{\rm q}(t')/2 \Big) \Big]   . \label{S_Phi-TrLn}
\end{multline}
 (The prefactor $1/2$ in front of ${\rm Tr \ln}$ is due the Pfaffian, which appears after  the integration over the non-independent Grassmann fields in the Gor'kov-Nambu formalism.)
 The Keldysh rotation  from $\Phi(t_\pm)$  to the classical and quantum components, $\Phi_{\rm cl}(t)=\frac{1}{2}(\Phi(t_+)+\Phi(t_-))$ and  $\Phi_{\rm q}(t)= \Phi(t_+) -  \Phi(t_-)$,   is performed.  
In the quasi-classical approach we expand  (\ref{S_Phi-TrLn})  up to second order in $\Phi_{\rm cl}$ and $\Phi_{\rm q}$. This gives a dissipative action of the Caldeira-Leggett type~\cite{CL}: 
\begin{multline} 
	\mathcal{S}
	[\Phi] 
	=   \frac{1}{8}\int  \frac{\omega^2}{ E_{\rm c}} \Phi_{\rm q, -\omega}  \Phi_{\rm cl, \omega}  \frac{d\omega}{2\pi} +
	\frac{1}{8} \int \frac{d\omega}{2\pi}
	 \times \\ 
	 \begin{bmatrix}
	\Phi_{\rm cl, -\omega}  & \Phi_{\rm q, -\omega} 
	\end{bmatrix}
	 \begin{bmatrix}
	0 & 
	-i\omega Y^*_\omega \\ \\
	i\omega Y_\omega & 
	 i {\rm Re}[Y_\omega] \omega \coth\frac{\omega}{2 T}  
	\end{bmatrix}	
	\begin{bmatrix}
	\Phi_{\rm cl, \omega}  \\  \\ \Phi_{\rm q, \omega}
	\end{bmatrix}
\label{S_Phi}
\end{multline}
  (see  the Appendix~\ref{App-CL} for details of the derivation).
This action is represented in Keldysh $\boldsymbol\sigma$-space  where the  matrix possesses the  causality structure~\cite{Kamenev}. There are retarded and advanced parts of the current correlator in the off-diagonal of this matrix. The Keldysh term in the right bottom corner reproduces fluctuation-dissipation theorem in this methodology   (see Sec.~\ref{App-Diss} in the Appendix). We note that the second order  expansion of the logarithm produces a divergent term, which, as usual, appears in Caldeira-Leggett theory with linearized coupling between $\Phi$ and $I$. The diamagnetic counter-term in (\ref{S-0}) ($\propto \omega_c$) cancels this divergency  (see Sec.~\ref{App-Admittance} in the Appendix).  

\subsection{ Non-stationary scattering matrix}
 Alternatively to the use of the step-like gauge transform that extends infinitesimally 
 into the incoming and the outgoing Dirac modes and
 that results in $\mathcal{S}_{\rm int}$, one can embed the superconducting phase into a non-stationary scattering matrix $S_\Phi(t,t')$, similarly to \cite{Levitov1996,PhysRevLett.85.1294,Kamenev}. This scattering matrix 
incorporates the propagation in the Majorana edge channels and
relates the local Dirac fields near the respective Y-splittings, 
 $\check\Psi_{\rm out}(z_2+\epsilon,t)=\int\check {S}(t,t') \check\Psi_{\rm in}(z_1-\epsilon,t') dt'$. Here, $ \check S(t-t')=\check R^+ \check F(t-t') \check R $ and   $\check F(t)=v\int\frac{dk}{2\pi}\check F_k e^{-ivk t}={\rm diag}\{ (-1)^{n+1}\delta(t-\tau_1),\ \delta(t-\tau_2) \}$.
 After the gauge transformation the scattering matrix acquires the form
 \begin{multline}
\check{S}_\Phi(t,t') = e^{-\frac{i}{2} \check \tau_z ( \boldsymbol{\sigma_0}\Phi_{\rm c}(t) +\boldsymbol{\sigma_z}\Phi_{\rm q}(t)/2)} \check S(t-t') \times \\ \times e^{\frac{i}{2} \check \tau_z ( \boldsymbol{\sigma_0}\Phi_{\rm c}(t') +\boldsymbol{\sigma_z}\Phi_{\rm q}(t')/2)} \ , \ \{t,t'\}\in [-\infty, \infty] \ .
\end{multline}
Integrating now over the fermionic degrees of freedom in the spirit of Refs.~\cite{PhysRevLett.85.1294,Kamenev} one could obtain the effective action for 
the phase variable $\Phi$.

\section{Conclusions} We have analyzed the microwave dynamics of a Majorana interferometer with floating 
superconducting island in the linear response regime. We show that one can observe 
the propagation and interference of Majorana excitations in the two branches of the interferometer by measuring the spectrum of microwaves reflected by the system. This is an alternative to proposals dealing with the detection of current and noise in the Ohmic contacts (sources or drains) of the interferometers. The proposed technique 
could also be used in the time-resolved manner, i.e, by sending microwave pulses and observing the response delayed due to the finite propagation time and the interference of the Majorana excitations. 

\section{  Acknowledgements.} We thank I. Pop for insightful discussions. This research was financially supported by the DFG-RFBR Grant [No. MI 658/12-1, SH 81/6-1  (DFG) and No. 20-52-12034 (RFBR)].


%

 \onecolumngrid
   \newcounter{defcounter}
\setcounter{defcounter}{0}

 \renewcommand\theequation{A\arabic{equation}}    
\setcounter{equation}{0}  

  \section{Appendix} \label{App-CL}
 \subsection{Derivation of the dissipative part in the effective action} \label{App-Diss}
 \subsubsection{Quasi-classical   approximation}
 The action for the phase $\Phi(t)$  reads:
\begin{equation}   
\mathcal{S} [\Phi] =\int_{\mathcal{C}} \Big(\frac{1}{2} \frac{1}{2E_{\rm c}}  \Big[\dot\Phi(t)/2\Big]^2  -\frac{1}{8} \langle U\rangle \Phi^2(t)\Big) dt  + \mathcal{S}_{\rm diss}[\Phi]  
	 \ . \label{App-S_QuasiCl}
\end{equation}
The first  term describes the charging energy; $E_{\rm c} \equiv 1/2 C_0$ (in full units $E_{\rm c} \equiv e^2/2C_0$). The second term is the diamagnetic one that follows from the quadratic expansion in (\ref{S-int}). The third term is a dissipative action 
 \begin{equation}
	\mathcal{S}_{\rm diss}[\Phi]  =     
		- i\frac{1}{2}{\rm Tr  \ln}  
	\Big[ \check{\mathbf{1}}   { +}\frac{1}{2} \check{\mathbf {G}}_{p}(t{-}t')   \check J_{ p,k}  ({\boldsymbol \sigma_z}\Phi_{\rm cl}(t')+  {\boldsymbol \sigma_0}\Phi_{\rm q}(t')/2) \Big] \ .  \label{App-S_Phi-TrLn}
\end{equation}
It is given by logarithm of a Pfaffian: $\ln {\rm Pf}[G^{-1}+\frac{1}{2} J  \Phi]=\frac{1}{2}{\rm Tr}\ln[G^{-1}+\frac{1}{2} J  \Phi]$. The Pfaffian appears after the integration over the fermion fields. In the quasi-classical regime, the quadratic expansion is employed: 
\begin{equation}
\mathcal{S}_{\rm diss}[\Phi]	=   
		- i\frac{1}{4}{\rm Tr  }_{\sigma,\tau} \Big[ 
	\int dt \int \frac{dp}{2\pi}  \check{\mathbf {G}}_{p}(0)   \check J_{ p,p}  ({\boldsymbol \sigma_z}\Phi_{\rm cl}(t)+  {\boldsymbol \sigma_0}\Phi_{\rm q}(t)/2) \Big] +  \label{App-S_Phi-TrLn-1}
\end{equation}
\begin{equation}   
		+ i\frac{1}{16}{\rm Tr   }_{\sigma,\tau}  \Big[ 
	\int dtdt' \int \frac{dp dk}{(2\pi)^2}\check{\mathbf {G}}_{p}(t{-}t')   \check J_{ p,k}  ({\boldsymbol \sigma_z}\Phi_{\rm cl}(t')+  {\boldsymbol \sigma_0}\Phi_{\rm q}(t')/2))\check{\mathbf {G}}_{k}(t'{-}t)   \check J_{ k,p}  ({\boldsymbol \sigma_z}\Phi_{\rm cl}(t)+  {\boldsymbol \sigma_0}\Phi_{\rm q}(t)/2)) \Big]  + ... \ . \label{App-S_Phi-TrLn-2}
\end{equation}
We perform the usual Keldysh rotation to the 
classical and quantum components of $\Phi(t)$. These are given by $\Phi(t_+)=\Phi_{\rm cl}(t)+\Phi_{\rm q}(t)/2$ and $\Phi(t_-)=\Phi_{\rm cl}(t)-\Phi_{\rm q}(t)/2$.
The current matrix $\check J_{ p,k}= v\check \tau_z-v \check S_{p}^+  \check \tau_z  \check S_{k}$  
(cf.~Eq.~(\ref{I[Psi]})) can be represented as
  \begin{equation}
	\check J_{ p,k}= v \left (V_{ p,k}\check\tau_y +W_{ p,k} \check\tau_z\right) \ , \label{App-J}
\end{equation}
where 
  \begin{equation}
  V_{ p,k} =(-1)^{n+1}e^{-\frac{iv(k-p)}{2E_{\rm Th}}} \sin\left( \frac{v (p+k)}{2\Lambda}\right)\ , \ W_{ p,k} = 1+(-1)^{n+1}e^{-\frac{iv(k-p)}{2E_{\rm Th}}} \cos\left( \frac{v (p+k)}{2\Lambda}\right) \ . \label{App-J-2}
 \end{equation}
 Above we have used the matrix valued Green's functions $\check{\mathbf {G}}$
 (the ``check'' symbol denotes the matrix structure in the Nambu space). These are given by
  \begin{equation}\check{\mathbf {G}}_{\omega,k} =\check\tau_+\check\tau_- \mathbf {G}_{\omega,k} - \check\tau_-\check\tau_+\mathbf {G}^T_{-\omega,-k}  \ . \end{equation}
 Here  ${\mathbf {G}}_{\omega,k}$ is a matrix in the Keldysh space given by 
  \begin{equation}
  {\mathbf {G}}_{\omega,k} =
  \begin{bmatrix} \frac{1}{\sqrt 2} & \frac{1}{\sqrt 2} \\ \\ 
		\frac{-1}{\sqrt 2} & \frac{1}{\sqrt 2}
	\end{bmatrix}
	\begin{bmatrix} g^{R}_{\omega,k} & g^{K}_{\omega,k}  \\ \\ 
		0 & g^{A}_{\omega,k} 
			\end{bmatrix}
	\begin{bmatrix} \frac{1}{\sqrt 2} & \frac{1}{\sqrt 2} \\ \\ 
		\frac{1}{\sqrt 2} & \frac{-1}{\sqrt 2}
	\end{bmatrix}
  \end{equation}
The unitary  matrices on the left and right hand sides produce the fermionic Keldysh rotation of the Green's function matrix in the middle. The latter is expressed in terms of the retarded/advanced functions, $g^{R/A}_{\omega,k}=(\omega-vk\pm i o)$, and Keldysh function, $g^K_{\omega,k}=(g^{R}_{\omega,k} - g^{A}_{\omega,k})f_k$. Here, $f_k=1-2n_k$ and $n_k=(1+\exp((vk-\Delta\mu)/T)^{-1}$ is the Fermi distribution function of the incident Dirac mode.  The temperature $T$ is determined by the Ohmic contact. $\Delta\mu$ is a difference between chemical potentials in the superconductor and the Ohmic contact.
 We study the equilibrium regime,  i.e.  $\Delta\mu=0$. In this case, $f_k=-f_{-k}=\tanh\frac{vk}{2T}$ and the Green's functions satisfy $\mathbf {G}^T_{-\omega,-k}=-\mathbf {G}_{\omega,k}$. Hence, for the Nambu matrix we obtain
 \begin{equation}\check{\mathbf {G}}_{\omega,k} =\check\tau_0 \mathbf {G}_{\omega,k}  \ . \end{equation}

  \subsubsection{Calculation of the trace over Nambu and Keldysh indices in $\mathcal{S}_{\rm diss}$}
 One can see that the first order term given by  (\ref{App-S_Phi-TrLn-1}) is zero  due to ${\rm Tr}_\tau [\check{\mathbf {G}}_{\omega,k}\check J_{ p,p}]=0$ in the equilibrium case. Let us now calculate  the second order term  (\ref{App-S_Phi-TrLn-2}). First, we calculate the trace over Nambu $\tau$-space (``check'' symbol is eliminated here):
 \begin{multline}   
	\mathcal{S}_{\rm diss}[\Phi]  = i\frac{1}{16}{\rm Tr   }_{\sigma}  \Big[ 
	\int dtdt' \int \frac{dp dk}{(2\pi)^2} 2v^2 (V_{ p,k}V_{ k,p}+W_{ p,k}W_{ k,p})\times\\ 
	\times {\mathbf {G}}_{p}(t{-}t')     ({\boldsymbol \sigma_z}\Phi_{\rm cl}(t')+  {\boldsymbol \sigma_0}\Phi_{\rm q}(t')/2)) {\mathbf {G}}_{k}(t'{-}t)      ({\boldsymbol \sigma_z}\Phi_{\rm cl}(t)+  {\boldsymbol \sigma_0}\Phi_{\rm q}(t)/2)) \Big]  \ . \label{App-S_Phi-tau}
\end{multline}
 Second, we use the Fourier transform, $\mathbf {G}_{p}(t)=\int \frac{d\omega}{2\pi} \mathbf {G}_{\omega,p}e^{-i\omega t}$, and transform the frequency variables as $\omega_1=\Omega+\omega/2$, $\omega_2  = \Omega - \omega/2$:
 \begin{multline}   
	\mathcal{S}_{\rm diss}[\Phi]  = i\frac{1}{16}
	\int dtdt' \int \frac{dp dk}{(2\pi)^2} 2v^2 (V_{ p,k}V_{ k,p}+W_{ p,k}W_{ k,p}) \int \frac{d\omega_1 d\omega_2}{(2\pi)^2} e^{-i(t-t')(\omega_1-\omega_2)} \times\\ 
	\times {\rm Tr   }_{\sigma}  \Big[ {\mathbf {G}}_{\omega_1,p}     ({\boldsymbol \sigma_z}\Phi_{\rm cl}(t')+  {\boldsymbol \sigma_0}\Phi_{\rm q}(t')/2)) {\mathbf {G}}_{\omega_2,k}      ({\boldsymbol \sigma_z}\Phi_{\rm cl}(t)+  {\boldsymbol \sigma_0}\Phi_{\rm q}(t)/2)) \Big] = \\
	= i\frac{1}{16}  
	\int dtdt' \int \frac{dp dk}{(2\pi)^2} 2 v^2(V_{ p,k}V_{ k,p}+W_{ p,k}W_{ k,p}) \int \frac{d\omega}{2\pi } e^{-i(t-t')\omega} \times\\ 
	\times \int \frac{  d\Omega}{2\pi }{\rm Tr   }_{\sigma}  \Big[ {\mathbf {G}}_{\Omega+\omega/2,p}     ({\boldsymbol \sigma_z}\Phi_{\rm cl}(t')+  {\boldsymbol \sigma_0}\Phi_{\rm q}(t')/2)) {\mathbf {G}}_{\Omega-\omega/2,k}      ({\boldsymbol \sigma_z}\Phi_{\rm cl}(t)+  {\boldsymbol \sigma_0}\Phi_{\rm q}(t)/2)) \Big]
	 \ . \label{App-S_Phi-tau-1}
\end{multline}
  Now we consider the last line in  (\ref{App-S_Phi-tau-1}) and calculate the trace over the Keldysh space $\sigma$. The result is expressed in terms of $g^{R/A}$ and $f_k$:
  \begin{multline}   
\int \frac{  d\Omega}{2\pi } {\rm Tr   }_{\sigma}  \Big[ {\mathbf {G}}_{\Omega+\omega/2,p}     ({\boldsymbol \sigma_z}\Phi_{\rm cl}(t')+  {\boldsymbol \sigma_0}\Phi_{\rm q}(t')/2)) {\mathbf {G}}_{\Omega-\omega/2,k}      ({\boldsymbol \sigma_z}\Phi_{\rm cl}(t)+  {\boldsymbol \sigma_0}\Phi_{\rm q}(t)/2)) \Big] = \\
=  
\int \frac{  d\Omega}{2\pi } \frac{1}{4}   g^A_{\Omega -\frac{\omega }{2},k} g^A_{\frac{\omega }{2}+\Omega ,p} (2 \Phi_{\rm cl}(t')-f_k \Phi_{\rm q}(t')) (2 \Phi_{\rm c}(t)-f_p \Phi_{\rm q}(t))
+\\
+\int \frac{  d\Omega}{2\pi } \frac{1}{4}   g^R_{\Omega -\frac{\omega }{2},k} g^R_{\frac{\omega }{2}+\Omega ,p} (f_k \Phi_{\rm q}(t)+2 \Phi_{\rm c}(t)) (f_p \Phi_{\rm q}(t')+2 \Phi_{\rm cl}(t'))+\\
+\int \frac{  d\Omega}{2\pi }  \frac{1}{4}   \left(2 \Phi_{\rm q}(t) \Phi_{\rm cl}(t') (f_p-f_k) g^A_{\Omega -\frac{\omega }{2},k} g^R_{\frac{\omega }{2}+\Omega ,p}  
+  2   \Phi_{\rm c}(t) \Phi_{\rm q}(t')(f_k-f_p) g^A_{\frac{\omega }{2}+\Omega ,p} g^R_{\Omega -\frac{\omega }{2},k}- \right. \\ \left.
- \Phi_{\rm q}(t)\Phi_{\rm q}(t') (f_k f_p-1) \left(g^A_{\Omega -\frac{\omega }{2},k} g^R_{\frac{\omega }{2}+\Omega ,p}+g^A_{\frac{\omega }{2}+\Omega ,p} g^R_{\Omega -\frac{\omega }{2},k}\right)\right) 
	 \ . \label{App-S_Phi-tau-2}
\end{multline}

\subsubsection{Integration over frequency in $\mathcal{S}_{\rm diss}$}
  The integration over $\Omega$ gives zero for the first and second integral in (\ref{App-S_Phi-tau-2}): $\int g^A_{\Omega -\frac{\omega }{2},k} g^A_{\frac{\omega }{2}+\Omega ,p} d\Omega =0$ and $\int  g^R_{\Omega -\frac{\omega }{2},k} g^R_{\frac{\omega }{2}+\Omega ,p} d\Omega =0$. This follows from the analytical properties of retarded and advanced Green functions. An integration of  cross terms in the last two lines in (\ref{App-S_Phi-tau-2}) with $g^{R/A}_{\omega,k}=(\omega-vk\pm i o)$ gives:
  \begin{equation}
 \int \frac{  d\Omega}{2\pi } g^A_{\Omega -\frac{\omega }{2},k} g^R_{\frac{\omega }{2}+\Omega ,p}  = i g^R_{ \omega   ,p-k}  \ , \ \int \frac{  d\Omega}{2\pi } g^A_{\frac{\omega }{2}+\Omega ,p} g^R_{\Omega -\frac{\omega }{2},k} = - i g^A_{ \omega   ,p-k} \ , 
  \end{equation} 
  and 
    \begin{equation}\int \frac{  d\Omega}{2\pi }\left(g^A_{\Omega -\frac{\omega }{2},k} g^R_{\frac{\omega }{2}+\Omega ,p}+g^A_{\frac{\omega }{2}+\Omega ,p} g^R_{\Omega -\frac{\omega }{2},k}\right)= i (g^R_{ \omega   ,p-k} -g^A_{ \omega   ,p-k}) = 2\pi \delta(\omega-v(p-k))\ .
  \end{equation}
  Finally, we have
   \begin{multline}   
\int \frac{  d\Omega}{2\pi } {\rm Tr   }_{\sigma}  \Big[ {\mathbf {G}}_{\Omega+\omega/2,p}     ({\boldsymbol \sigma_z}\Phi_{\rm cl}(t')+  {\boldsymbol \sigma_0}\Phi_{\rm q}(t')/2)) {\mathbf {G}}_{\Omega-\omega/2,k}      ({\boldsymbol \sigma_z}\Phi_{\rm cl}(t)+  {\boldsymbol \sigma_0}\Phi_{\rm q}(t)/2)) \Big] = \\
=  
 \frac{1}{2} \left( i \Phi_{\rm q}(t) \Phi_{\rm cl}(t') (f_p-f_k)  g^R_{ \omega   ,p-k}  
-  i   \Phi_{\rm c}(t) \Phi_{\rm q}(t')(f_k-f_p) g^A_{ \omega   ,p-k}  
- \Phi_{\rm q}(t)\Phi_{\rm q}(t') (f_k f_p-1) \pi \delta(\omega-v(p-k)) \right) 
	 \ . \label{App-S_Phi-tau-3}
\end{multline}
 At this stage we embed the result (\ref{App-S_Phi-tau-3}) into (\ref{App-S_Phi-tau-1}) and obtain
  \begin{multline}   
	\mathcal{S}_{\rm diss}[\Phi]  = \frac{1}{16}  
	\int dtdt' \int \frac{dp dk}{(2\pi)^2}  v^2 (V_{ p,k}V_{ k,p}+W_{ p,k}W_{ k,p}) \int \frac{d\omega}{2\pi } e^{-i(t-t')\omega} \times\\ 
	\times    \left( \Phi_{\rm q}(t) \Phi_{\rm cl}(t') (f_k-f_p)  g^R_{ \omega   ,p-k}  
+  \Phi_{\rm c}(t) \Phi_{\rm q}(t')(f_k-f_p) g^A_{ \omega   ,p-k}   
- i \Phi_{\rm q}(t)\Phi_{\rm q}(t') (f_k f_p-1) \pi \delta(\omega-v(p-k)) \right) 
	 \ . \label{App-S_Phi-tau-4} 
\end{multline}
\subsubsection{Calculation of the retarded part in $\mathcal{S}_{\rm diss}$}
After the Fourier transform in (\ref{App-S_Phi-tau-4}) for the fields, $\Phi_{\rm cl/q}(t)=\int \frac{d\omega}{2\pi} \Phi_{\rm cl/q, \omega}$, we represent  the action in the standard Keldysh form:
 \begin{equation}   
\mathcal{S}_{\rm diss}[\Phi]  =\frac{1}{8}\int \frac{d\omega}{2\pi }  
	   \begin{bmatrix}
	\Phi_{\rm cl, -\omega}  & \Phi_{\rm q, -\omega} 
	\end{bmatrix}	
	\begin{bmatrix}
0 &   \mathcal{L}_\omega^* \\ \\
\mathcal{L}_\omega & \mathcal{K}_\omega
\end{bmatrix}
\begin{bmatrix}
	\Phi_{\rm cl, \omega}  \\  \\ \Phi_{\rm q, \omega}
	\end{bmatrix}
	 \ . \label{App-S_Phi-tau-5}
\end{equation}
Here, the function $\mathcal{L}_\omega$ and its complex conjugate $\mathcal{L}_\omega^*$ are the retarded and advanced current-current correlators, respectively, whereas $\mathcal{K}_\omega$ is the Keldysh component. 

We now calculate the retarded component $\mathcal{L}_\omega$:
\begin{equation}
\mathcal{L}_\omega=\int \frac{dp dk}{(2\pi)^2}   \frac{ v^2}{2} (V_{ p,k}V_{ k,p}+W_{ p,k}W_{ k,p})(f_k-f_p)  g^R_{ \omega   ,p-k}  \ .
\end{equation}
We start the calculation of this integral with the use of (\ref{App-J}, \ref{App-J-2}) for the current-current term that reads \begin{equation}(V_{ p,k}V_{ k,p}+W_{ p,k}W_{ k,p})=2-2(-1)^n\cos(\varepsilon/(2E_{\rm Th}))\cos(\xi/\Lambda) \ . \end{equation} We change momentum variables as  $k=v^{-1}(\xi -\varepsilon/2)$ and $p=v^{-1}(\xi +\varepsilon/2)$ where $\xi$ and $\varepsilon$ are new frequencies.  
 After this transform, the 
  integration over $\xi$ is performed as follows:
 \begin{equation}
\mathcal{L}_\omega=\int \frac{  d\varepsilon}{(2\pi)^2}    \frac{1}{\omega-\varepsilon+io} \int  d\xi   \Big[1-(-1)^n\cos(\varepsilon/(2E_{\rm Th}))\cos(\xi/\Lambda) \Big]\left(\tanh\frac{\xi -\varepsilon/2}{2T}-\tanh\frac{\xi +\varepsilon/2}{2T}\right)   =
\end{equation} 
   \begin{equation}
=\int \frac{  d\varepsilon}{(2\pi)^2}    \frac{2}{\omega-\varepsilon+io}    \Big[- \varepsilon +2 (-1)^n \frac{\pi T }{\sinh\frac{\pi T}{\Lambda}}\cos(\varepsilon/(2E_{\rm Th}))\sin(\varepsilon/(2\Lambda)) \Big] \ . 
\end{equation} 
   Before we perform the last integration over $\varepsilon$, we extract the linear divergent constant term.  We also use the representation $E_{\rm Th}=\frac{v}{l_1 +l_2}$ and $\Lambda=\frac{v}{l_2 -l_1} $ assuming that $l_2>l_1$:
   \begin{equation}
\mathcal{L}_\omega=\frac{1}{2\pi^2}\int\limits_{-\omega_c}^{\omega_c}d\varepsilon + \int \frac{  d\varepsilon}{2\pi^2}    \frac{1}{\varepsilon-\omega-io}    \Big[\omega +\frac{i}{2}(-1)^n \frac{\pi T }{\sinh\frac{\pi T}{\Lambda}}\Big( e^{-i\frac{l_1\varepsilon}{v}} - e^{i\frac{l_1\varepsilon}{v}} - e^{-i\frac{l_2\varepsilon}{v}} + e^{i\frac{l_2\varepsilon}{v}}\big) \Big] = 
\end{equation} 
   \begin{equation}
=\frac{ \omega_c}{\pi^2} +   \frac{ 1}{2\pi^2}       \Big[i\pi \omega +2\pi i \frac{i}{2}(-1)^n \frac{\pi T }{\sinh\frac{\pi T}{\Lambda}}\big( e^{i\frac{l_2\omega}{v}} - e^{i\frac{l_1\omega}{v}} \big) \Big] \ . 
\end{equation} 
Thus, we arrive at one of the central results of this work:
   \begin{equation}
\mathcal{L}_\omega=\frac{ \omega_c}{\pi^2} +   \frac{    i  \omega}{2 \pi }       \Big[1 + (-1)^n \frac{\pi T }{\sinh\frac{\pi T}{\Lambda}}\frac{\big( e^{i\frac{l_1\omega}{v}} - e^{i\frac{l_2\omega}{v}} \big)}{i\omega} \Big] \ . \label{App-L}
\end{equation} 
We note  that the presence of the imaginary and linear in frequency term, $\propto i \omega$,  is accompanied by   the divergent  real one, $\propto \omega_c$. This is dictated by the  analytical (Kramers-Kronig) structure of the retarded function $\mathcal{L}_\omega$.

\subsubsection{Calculation of the Keldysh part in $\mathcal{S}_{\rm diss}$. Fluctuation-dissipation relation}

The Keldysh component is given by 
\begin{equation}
\mathcal{K}_\omega=i\pi \frac{v^2}{2} \int \frac{dp dk}{(2\pi)^2} (V_{ p,k}V_{ k,p}+W_{ p,k}W_{ k,p})\Big(1-f_k f_p\Big)\delta(\omega-v(p-k)) =   
\end{equation}
\begin{equation}
 =i\pi   \int \frac{d\varepsilon d\xi}{(2\pi)^2} (1- (-1)^n\cos(\varepsilon/(2E_{\rm Th}))\cos(\xi/\Lambda))\Big(1-\tanh\frac{\xi -\varepsilon/2}{2T}\tanh\frac{\xi +\varepsilon/2}{2T}\Big)   \delta(\omega-\varepsilon)= 
\end{equation}
\begin{equation}
 =i\pi   \int \frac{ d\xi}{(2\pi)^2} (1- (-1)^n\cos(\omega/(2E_{\rm Th}))\cos(\xi/\Lambda))\Big(1-\tanh\frac{\xi -\omega/2}{2T}\tanh\frac{\xi +\omega/2}{2T}\Big)   = 
\end{equation}
\begin{equation}
 = \frac{ i}{ 2\pi}  \Big(\omega- 2(-1)^n\frac{\pi T}{ \sinh\frac{\pi T}{\Lambda}} \cos(\omega/(2E_{\rm Th}))\sin(\omega/(2\Lambda))\Big)  \coth \frac{\omega}{2T} \  .  
\end{equation}
We note that using (\ref{App-L}) we obtain the relation
\begin{equation}
\mathcal{K}_\omega = i {\rm Im} [\mathcal{L}_\omega]\coth \frac{\omega}{2T}   \ , \label{App-K}
\end{equation}
which reflects the fluctuation-dissipation theorem.

\subsection{Cancellation of the divergent counter term. Derivation of the admittance $Y_\omega$}\label{App-Admittance}
\subsubsection{Green functions for the phase}
The quasi-classical action (\ref{App-S_QuasiCl}) now reads:
\begin{multline}   
\mathcal{S} [\Phi] =\int_{\mathcal{C}} \Big(\frac{1}{2} \frac{1}{2E_{\rm c}}  \Big[\dot\Phi(t)/2\Big]^2  -\frac{1}{8} \langle U\rangle \Phi^2(t)\Big) dt  + \mathcal{S}_{\rm diss}[\Phi]  = \\= \int \Phi_{\rm q}(t)\left(-\frac{\langle U\rangle}{4}-\frac{1}{ 
		8E_{\rm c}}\partial_t^2\right)\Phi_{\rm cl} (t)dt  + \mathcal{S}_{\rm diss}[\Phi]  = \\=\frac{1}{8}\int \frac{d\omega}{2\pi }  
	   \begin{bmatrix}
	\Phi_{\rm cl, -\omega}  & \Phi_{\rm q, -\omega} 
	\end{bmatrix}	
	\begin{bmatrix}
0 &   \frac{\omega^2}{2E_{\rm c}} - \langle U\rangle+ \mathcal{L}_\omega^* \\ \\
\frac{\omega^2}{2E_{\rm c}} - \langle U\rangle+\mathcal{L}_\omega & \mathcal{K}_\omega
\end{bmatrix}
\begin{bmatrix}
	\Phi_{\rm cl, \omega}  \\  \\ \Phi_{\rm q, \omega}
	\end{bmatrix}
	 \ . \label{App-S_Phi-tau-FINAL}
\end{multline}
Since $\Phi$ is real, the action (\ref{App-S_Phi-tau-5}) can be represented as 
\begin{equation}\mathcal{S} [\Phi] =\frac{1}{2}\int\frac{d\omega}{2\pi } \sum\limits_{\sigma,\sigma'={\rm cl,q}}\Phi_{\sigma,-\omega} [\mathbf{D}^{-1}_\omega]_{\sigma, \sigma'}\Phi_{\sigma',\omega} \ ,
\end{equation}
where 
\begin{equation} \mathbf{D}_\omega= \begin{bmatrix}
 D_\omega^K & D_\omega^R\\ \\
D_\omega^A   & 0
\end{bmatrix} 
\end{equation}
possesses the usual Keldysh structure:
\begin{equation}
\ D_\omega^R= \frac{4}{ \frac{\omega^2}{2E_{\rm c}} - \langle U\rangle+ \mathcal{L}_\omega}  \ , \ D_\omega^A= \frac{4}{ \frac{\omega^2}{2E_{\rm c}} - \langle U\rangle+ \mathcal{L}_\omega^*} 
 \  , \ D_\omega^K = (D^R_\omega-D^A_\omega)\coth \frac{\omega}{2T}
 \ .
 \end{equation}
 The fluctuation-dissipation relation (\ref{App-K}) for $\mathcal{K}_\omega$ has been used in $D^K_\omega$.
 
 \subsubsection{Analysis of the linear response function. Divergent terms cancellation. Admittance}
 The retarded component $D^R_\omega$ determines the response function for the phase variable. If an external current perturbation $I_{\rm ext}(t)$ is applied then the following perturbation is added to the action $S_{\rm ext}= -\int  \hbar^{-1} H_{\rm ext} dt=\int \frac{1}{2e}\Phi(t) I_{\rm ext}(t)dt $. The response   is $\Phi_\omega=\frac{1}{2}D^R_\omega I_{\rm ext, \omega}$.
 Assuming that the induced voltage $V_{\rm ind}$ is the difference between the potential in the Ohmic contact, which is set to zero, and in the superconductor, $V_\omega=-i\omega \Phi_\omega/2$, we have \begin{equation}V_{\rm ind, \omega}= 0 - V_\omega= i\omega \Phi_\omega/2  \ . 
 \end{equation}
 Then, we obtain that
$V_{\rm ind, \omega}=  i\omega \frac{1}{4}D^R_\omega I_{\rm ext, \omega}  $.  The total admittance, $Y_{\rm tot, \omega}= \frac{I_{\rm ext, \omega}}{V_{\rm ind, \omega}}$, is given by 
  \begin{equation}Y_{\rm tot, \omega} =  \frac{-i \omega}{2 E_{\rm c}} +\frac{1}{i\omega}\left(\frac{ \omega_c}{\pi^2} - \langle U\rangle \right)+  Y_\omega    \ . 
 \end{equation}
 The first term is the  admittance of the capacitor $C_0$. The second term in braces consists of a divergent inductive part and a counter-term. As shown after Eq.~(\ref{S-0}) these two terms 
 cancel each other. The third term is the admittance of the system:
   \begin{equation}Y_\omega=  \left(\frac{e}{\hbar} \right) \frac{  1}{2 \pi }       \Big[1 + (-1)^n \frac{\pi T }{\sinh\frac{\pi T}{\Lambda}}\frac{\big( e^{i\frac{l_1\omega}{v}} - e^{i\frac{l_2\omega}{v}} \big)}{i\omega} \Big] \end{equation}
   (we added here the dimensional prefactor $e/\hbar$ and reinstate the conductance quantum $G_0$ in the Eq. (\ref{eq:Admittance})).
 Finally, we note that the retarded and Keldysh components in the action (\ref{App-S_Phi-tau-FINAL}) can be represented as follows:
  \begin{equation}\mathcal{L}_\omega= i \omega Y_\omega \ , \ \mathcal{L}_\omega^*= -i \omega Y_\omega^* \ , \  \mathcal{K}_\omega= i  {\rm Re}[Y_\omega] \omega \coth\frac{\omega}{2T}\ . \end{equation}
 Thus, one 
   arrives at Eq. (\ref{S_Phi}). 
   
\end{document}